\documentclass[twocolumn]{aastex63}

\usepackage{nicefrac}
\usepackage{amsmath}
\usepackage{float}
\usepackage{bm}
\usepackage{multirow}
\usepackage{xspace}

\usepackage{chngcntr}
\counterwithout{table}{section}
\counterwithout{table}{subsection}

\usepackage{tikz}
\usetikzlibrary{shapes.misc}

\tikzset{cross/.style={cross out, draw=black, minimum size=2*(#1-\pgflinewidth), inner sep=0pt, outer sep=0pt},
cross/.default={1pt}}

\newcommand{\ud}{\mathrm{d}}

\newcommand{\code}[1]{{\tt #1}}

\newcommand{\msun}{M_{\odot}}
\newcommand{\bhspin}{a_\star}
\newcommand{\mbh}{M}
\newcommand{\bE}{\mathbf{E}}
\newcommand{\bB}{\mathbf{B}}
\newcommand{\moscref}{M11\xspace}

\begin{document}

\title{Pair Drizzle around Sub-Eddington Supermassive Black Holes}

\correspondingauthor{George N. Wong}
\email{gnwong2@illinois.edu}

\author[0000-0001-6952-2147]{George~N.~Wong}
\affil{Illinois Center for Advanced Studies of the Universe, Department of Physics, \\University of Illinois, 1110 West Green Street, Urbana, IL 61801, USA}
\author[0000-0001-8939-4461]{Benjamin~R.~Ryan}
\affil{CCS-2, Los Alamos National Laboratory, P.O. Box 1663, Los Alamos, NM 87545, US}
\author[0000-0001-7451-8935]{Charles~F.~Gammie}
\affil{Illinois Center for Advanced Studies of the Universe, Department of Physics, \\University of Illinois, 1110 West Green Street, Urbana, IL 61801, USA}
\affil{Department of Astronomy, University of Illinois, 1002 West Green Street, Urbana, IL 61801, USA}

\begin{abstract}

Electron--positron pair creation near sub-Eddington accretion rate black holes is believed to be dominated by the Breit--Wheeler process (photon--photon collisions).  The interacting high energy photons are produced when unscreened electric fields accelerate leptons either in coherent, macroscopic gaps or in incoherent structures embedded in the turbulent plasma flow. The latter type of acceleration results in a drizzle of pair production sourced by photons from the background radiation field whose energies are near the pair-production threshold. In this work, we use radiation GRMHD simulations to extend an earlier study of {\emph{pair drizzle}} by Mo\'scibrodzka et al.~We focus on low-magnetization (SANE) accretion onto supermassive Kerr black holes and consider radiation due to synchrotron, bremsstrahlung, and Compton upscattering processes. We confirm that pair drizzle in M87 is sufficient to keep the magnetospheric charge density orders of magnitude above the Goldreich--Julian density.  We also find that pair production peaks along the jet--disk boundary.

\end{abstract}


\section{Introduction}

The Event Horizon Telescope (EHT) recently published the first resolved images of plasma surrounding the M87 black hole at 1.3mm \citep{PaperI,PaperIV}.
Although these images and anticipated future results carry information about physical conditions in the accreting plasma, an accurate model of the emission source---the radiating leptons---must be obtained in order to extract the information \citep[see][]{PaperV}.  The subset of the radiating leptons that originates as electron--positron pairs is of particular interest (see \citealt{svensson89} for a review), and with this goal in mind, we consider a nearly {\em ab initio} model of pair production in low accretion rate (highly sub-Eddington) systems like M87 \citep[see][]{Broderick2015m87leptons,hirotani2018highenergyemission}.

Models of pair production around low accretion rate black holes require a population of high energy leptons that can Compton upscatter the low frequency background photons produced by the hot plasma. The upscattered high energy photons produce electron--positron pairs through interactions with the fiducial low-energy background photons via the \citet{breit34} process when the center-of-momentum energy of the interacting photons exceeds the rest-mass energy of an electron--positron pair $\sim$~1~MeV.

High-energy leptons can be produced in a variety of ways.
Gap models envisage coherent regions with $\bE \cdot \bB \ne 0$ that accelerate the leptons and initiate pair cascades \cite{beskin92, hirotani98, medvedev17, levinson18, chen18, Parfrey2019}. In gap models, the high-energy photons typically have energies that are orders of magnitude above the MeV threshold.
In contrast, {\emph{drizzle}} models predict that the native high energy component of the electron distribution throughout the near-horizon plasma will produce a steady, smooth background of $\sim$~MeV photons that interact with each other and pair produce (see \citealt{moscibrodzka11}, hereafter \moscref, and also \citealt{levinson18}).
Although gap and drizzle models may appear distinct, they can be thought of as end members of a continuum in which the structures that accelerate the leptons range from  coherent, steady, and large scale (gap) to incoherent, transient, and small scale (drizzle).

In this paper we revisit the drizzle model of \moscref, which estimated pair production rates based on nonradative general relativistic magnetohydrodynamics (GRMHD) accretion simulations.
We extend the \moscref estimate using {\em radiative} GRMHD (radGRMHD) accretion simulations produced with the \code{ebhlight} code \citep{ryan15,ebhlight}. \code{ebhlight} independently tracks the ion and electron temperatures~\citep{ressler15,ryan17a} and implements a
more thorough treatment of electron thermodynamics that explicitly includes both a model to partition dissipation between electrons and ions and a treatment of ion--electron energy exchange through Coulomb scattering \citep{ressler15}.  Moreover, \code{ebhlight} accurately accounts for radiative cooling by solving the radiation transport equation with a Monte Carlo method, which can be important to the plasma dynamics as accretion rates increase.  Furthermore, in contrast to the \moscref model, our pair production calculation includes photons produced by bremsstrahlung emission, which are unimportant for the thermal evolution of the fluid at the accretion rates we consider but may play an important role in drizzle pair production due to their characteristic high frequencies \citep{yarza2020bremsstrahlung}.  These extensions improve the accuracy of the pair production rate evaluation, especially at high accretion rates.

It is computationally expensive both to produce radGRMHD simulations and to generate well-resolved samples of the radiation field, so we evaluate pair drizzle for a targeted set of axisymmetric models.  We consider two black hole spins $\bhspin \equiv Jc/G\mbh^2 = 0.5$ and $0.94$ (here $J$ and $\mbh$ are the angular momentum and mass of the black hole respectively) over a range of mass accretion rates $\dot{m}\equiv \dot{M}/ \dot{M}_{\mathrm{Edd}}$\footnote{Here, $\dot{M}_{\mathrm{Edd}} \equiv L_{\mathrm{Edd}} / \left( \eta \,c^2 \right)$ where $\eta = 0.1$ is the reference nominal accretion efficiency, and the Eddington luminosity $L_{\mathrm{Edd}} \equiv 4 \pi G M m_p c / \sigma_T$} corresponding to geometrically thick, optically thin, weakly radiative accretion flows in the low magnetic flux ``standard and normal evolution'' (SANE) accretion state.\footnote{SANE in contrast to ``magnetically arrested disk'' (MAD) models, which have magnetic flux through the event horizon $\Phi$ satisfying $\phi \equiv \Phi/(\dot{M} r_g^2 c)^{1/2} \simeq 15$; here $r_g = G M/c^2$.}

The paper is organized as follows. Section~\hyperref[sec:models]{2} reviews the governing equations of radGRMHD, and Section~\hyperref[sec:pairs]{3} describes the pair production model and its implementation. In Section~\hyperref[sec:spatial]{4}, we discuss the expected spatial dependence of drizzle pair production. Section~\hyperref[sec:results]{5} presents the results of our numerical simulations. We discuss some physical implications and model limitations in Section~\hyperref[sec:discussion]{6}, and we provide a summary in Section~\hyperref[sec:summary]{7}.

\section{Plasma Model}
\label{sec:models}

We consider prograde black hole accretion, in which the orbital angular momentum of the plasma is aligned with the spin of the central black hole. Our models have accretion rates $\dot{m} \le 10^{-5}$. We find that these accretion rates are low enough for the plasma to be $\sim$ collisionless (i.e., the Coulomb scattering mean free path for electrons and ions is large compared to $G M/c^2$) but high enough that radiative cooling may influence the electron temperature. Hereafter, we set $G M = c = m_e = 1$ and occasionally restore cgs units for clarity.  
We model the plasma using radiative general relativistic magnetohydrodynamics (radGRMHD). 

In a coordinate basis, the governing equations of radGRMHD are
\begin{align}
\partial_t \left( \sqrt{-g} \rho_0 u^t \right) &= -\partial_i \left( \sqrt{-g} \rho_0 u^i \right), \label{eqn:massConservation}\\
\label{eqn:stressEnergyConservation}
    \partial_t \left( \sqrt{-g} {T^t}_{\nu} \right) &={} \partial_i \left( \sqrt{-g} {T^i}_{\nu} \right) + \sqrt{-g} {T^{\kappa}}_{\lambda} {\Gamma^{\lambda}}_{\nu\kappa} \nonumber \\
& \quad - \sqrt{-g}{R^{\mu}}_{\nu;\mu},
\\
\partial_t \left( \sqrt{-g} B^i \right) &= \partial_j \left[ \sqrt{-g} \left( b^j u^i - b^i u^j \right) \right], \label{eqn:fluxConservation} \\
\partial_i \left( \sqrt{-g} B^i \right) &= 0, \label{eqn:monopoleConstraint} 
\end{align}
where the plasma is defined by its rest mass density $\rho_0$, its four-velocity $u^\mu$, and $b^\mu$ is the magnetic field four-vector following \citet{mckinney04}. Here, $g \equiv {\rm det}(g_{\mu\nu})$ is the determinant of the covariant metric, $\Gamma$ is a Christoffel symbol, and $i$ and $j$ denote spatial coordinates.
In Equations~\ref{eqn:fluxConservation} and~\ref{eqn:monopoleConstraint}, we express components of the electromagnetic field tensor $F^{\mu\nu}$ as $B^i \equiv \star F^{it}$ for notational simplicity.

The stress-energy tensor ${T^\mu}_\nu$ contains contributions from both the fluid and the electromagnetic field:
\begin{align}
T^{\mu}_{\nu} &= \left( \rho_0 + u + P + b^{\lambda}b_{\lambda}\right)u^{\mu}u_{\nu} \nonumber \\
& \quad + \left(P + \frac{b^{\lambda}b_{\lambda}}{2} \right)g^{\mu}_{\nu} - b^{\mu}b_{\nu},
\end{align}
where $u$ is the internal energy of the fluid and the fluid pressure $P$ is related to its internal energy through an adiabatic index $\hat{\gamma}$ with $P \equiv \left(\hat{\gamma} - 1\right) u$ (see, e.g., \citealt{mckinney03}).

The radiation stress tensor is 
\begin{align}
{R^{\alpha}}_{\beta} = \int \frac{d^3 p}{\sqrt{-g}p^t}p^{\alpha}p_{\beta} \left(\frac{I_{\nu}}{h^4 \nu^3}\right),
\end{align}
where $p^\alpha$ is the four-momentum of a photon, $\nu$ is the frequency of the photon, and $I_\nu$ is specific intensity. Photons obey the equations of radiative transfer equations as they move through the plasma:
\begin{align}
\frac{dx^{\alpha}}{d\lambda} &= k^{\alpha}, \label{eqn:photonMotion} \\
\frac{dk^{\alpha}}{d\lambda} &= -\Gamma^{\lambda}_{~\alpha\beta}k^{\alpha}k^{\beta}, \label{eqn:geodesic} \\
\frac{D}{d\lambda}\left(\frac{I_{\nu}}{\nu^3}\right) &= \frac{\eta_{\nu}(T_e)}{\nu^2} -\frac{I_{\nu}\chi_{\nu}(T_e)}{\nu^2}. \label{eqn:radiativeTransfer} 
\end{align}
Here, $\eta_\nu$ is the local emissivity of the plasma, and $\chi_\nu$ encodes the total (scattering and absorption) opacity due to thermal synchrotron processes and Compton scattering.

We consider a two temperature plasma composed of electrons and ions. The extra degree of freedom is closed through an independent electron energy equation as in \citet{ressler15}:
\begin{align}
    \frac{\rho^{\hat{\gamma}_e}}{\hat{\gamma}_e - 1} u^{\mu} \partial_{\mu} \kappa_e &= f_e Q_H + Q_C(T_e, T_p) - u^{\nu}{R^{\mu}}_{\nu;\mu}, \label{eqn:electronEntropy}
\end{align}
where $\hat{\gamma}_e$ is the adiabatic index of the electrons, $f_e$ is the fraction of the volumetric dissipation rate $Q_H$ that goes into electrons (taken from the \citealt{howes10} model), and $Q_C$ is volumetric heating or cooling due to Coulomb scattering \citep{stepney83}.

\subsection{Two-Temperature Radiative General Relativistic Magnetohydrodynamics}

We solve the governing equations using the \code{ehblight} code \citep{ryan17a}. In \code{ebhlight}, the radiation stress-energy tensor is co-evolved with the fluid and is computed at each step from Monte Carlo samples of the radiation field, which are evolved according to the scheme introduced in \code{grmonty} \citep{dolence09}. \code{ebhlight} also independently tracks the proton and electron temperatures according to a two-temperature model where the electron entropy is evolved as in \cite{ressler15}.

\section{Pair Production}
\label{sec:pairs}

Pair drizzle in low accretion rate systems is weak, so the radiation field can be treated as independent of pair production. The pair production rate can thus be evaluated in a post-processing step after the fluid evolution has been completed.  We will show below that this approximation is self-consistent. 

\subsection{Comparison of Contributing Interactions}

The pair production rate density is 
\begin{equation}
\label{eq:general-interaction-eqn}
\dot{n}_\pm = n_1 \, n_2 \left< \sigma_{12} \, v \right>
\end{equation}
where $n_1$ and $n_2$ are the number densities of the two interacting species, $\sigma_{12}$ is their interaction cross section, $v$ is their relative velocity, and the angle brackets indicate an average over state variables. 

In our radiative electron--ion plasma, pair-producing interactions can occur between electrons ($e$), ions ($p$), and photons ($\gamma$).  The pair production cross sections are \citep[][]{stepney83, phinney83, zdziarski1985, phinney95, krolik99}
\begin{equation}
\sigma_{pp} \sim \sigma_{ee} \sim \sigma_{ep} \sim \alpha \sigma_{p\gamma} \sim \alpha \sigma_{e\gamma} \sim \alpha^2 \sigma_{\gamma\gamma},
\end{equation}
where $\alpha \approx 1/137$ is the fine-structure constant. Which process dominates depends on the details of the radiation field and plasma density.  We can estimate $n_e$ using {\tt ebhlight} simulations for guidance and assuming a pure hydrogen plasma; we can also estimate $n_\gamma$ by analyzing the simulated radiation field and counting only photons with energy $ > m_e c^2$. Then in the low density \emph{jet} region near the spin axis of the black hole, we find that $n_\gamma / n_{e} = n_\gamma / n_p > 1 > \alpha$.\footnote{\code{ebhlight} can resolve only a limited density contrast, so the density is artificially increased in the jet via numerical ``floors''.  The numerical electron density is therefore an upper limit on the physical density.} The $\gamma\gamma$ process therefore dominates pair production.

\subsection{Basic Equations}

The pair production rate density due to the $\gamma\gamma$ process (counting pairs and not individual particles) is
\begin{align}
\label{eq:ndot}
\dot{n}_\pm \equiv& \dfrac{1}{\sqrt{-g}}\dfrac{\ud N_\pm}{\ud^3 x \, \ud t} \nonumber \\
=& \dfrac{1}{2} \int \dfrac{\ud^3 k}{\sqrt{-g}} \dfrac{\ud^3 k'}{\sqrt{-g}} \dfrac{\ud N_\gamma}{\ud^3 x \, \ud^3 k} \dfrac{\ud N_\gamma}{\ud^3 x \, \ud^3 k'} \dfrac{\epsilon^2}{k^0 k'^0}\, \sigma_{\gamma\gamma} \, c,
\end{align}
where $\ud N_\gamma / \ud^3 x \, \ud^3 k$ is the photon distribution function, the factor of \nicefrac{1}{2} prevents double counting of interacting photons, and the center-of-momentum energy $\epsilon$ and cross section $\sigma_{\gamma\gamma}$ are 
\begin{align}
\epsilon^2 &= -\dfrac{k_\mu k'^\mu}{2}, \\
\sigma_{\gamma\gamma} &= \dfrac{3 \sigma_T}{8 \epsilon^6} \left[ \left( 2 \epsilon^4 + 2 \epsilon^2 - 1\right) \cosh^{-1} \epsilon - \vphantom{ \sqrt{\epsilon^2 - 1} } \right. \nonumber \\
& \qquad \left. \left( \epsilon^3 + \epsilon\right) \sqrt{\epsilon^2 - 1} \right],
\end{align}
where $\sigma_T$ is the Thomson cross section \citep[see][]{breit34}.  Note that the phase space volume element $\ud^3 x \, \ud^3 k$ is only invariant if the integration is over the components of the covariant wave four-vector, i.e., $\ud^3 k \equiv \ud k_1 \, \ud k_2 \, \ud k_3$. 

\begin{figure}[!ht] \begin{center}
\begin{tikzpicture}
\draw[thick,black] (-0.5,0) -- (3.5,0);
\draw[thick,black] (0,-2) -- (0,2);
\draw (0,1.8) node[cross=4pt,thick,red] {};
\draw (0,-1.8) node[cross=4pt,thick,red] {};
\draw[black] (0.1,1.95) node[right] {emitter};
\draw[black] (0.1,-1.95) node[right] {emitter};
\draw[thin,black] (0,1.8) -- (3,0);
\draw[thin,black] (0,-1.8) -- (3,0);
\draw[thin,<->] (-0.22,0.1) -- (-0.22,1.7);
\draw[black] (-0.22,0.9) node[left] {$L$};
\draw[thin,->] (0.1,0.15) -- (2.5,0.15);
\draw[black] (1.3,0.15) node[above] {$x$};
\draw[black] (2.7,0.2) arc (150:210:0.4);
\draw (2.8,0.24) node[above] {$\theta$};
\end{tikzpicture}
  \caption{The test problem geometry comprises two isotropic emitters separated by a distance $2L$. The pair production rate density $\dot{n}_\pm(x)$ is evaluated as a function of distance $x$ along the perpendicular bisector of the two sources. The angle between two incident photons at a point $x$ along the bisector is $\theta = 2\arctan(L/x) \approx 2L/x$ for $x/L \gg 1$.}
  \label{fig:twopoint-geometry} 
\end{center} \end{figure}
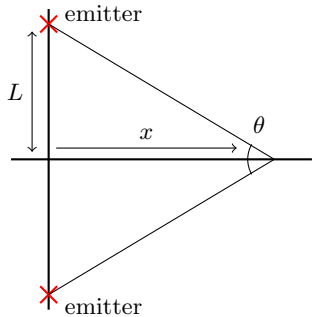

The photon distribution function (i.e., the radiation field) is generated by the same synchrotron emission and absorption plus Compton scattering physics of the radGRMHD model; however, for the pair computation we include an additional model for bremsstrahlung emission (bremsstrahlung absorption is negligible). We note that bremsstrahlung is energetically subdominant everywhere in our models, but it may be an important source of high energy (and therefore pair-producing) photons.
We adopt the piecewise bremsstrahlung emissivity of \cite{straub12} \citep[see also][]{yarza2020bremsstrahlung}.

\subsection{Numerical Implementation of Pair Production}

In radGRMHD, the plasma evolution depends on the radiation stress-energy tensor, which is an integral over the entire photon distribution function.  In contrast, the $\gamma\gamma$ pair production rate is a double integral over the photon distribution function and is dominated by a small range of energies around the pair production threshold.  The pair production rate calculation therefore requires a more accurate estimate of the photon distribution function than does the plasma evolution.  This is the main numerical motivation for evaluating the pair production in post-processing.

\begin{figure}[!ht] \begin{center}
  \includegraphics[width=\linewidth]{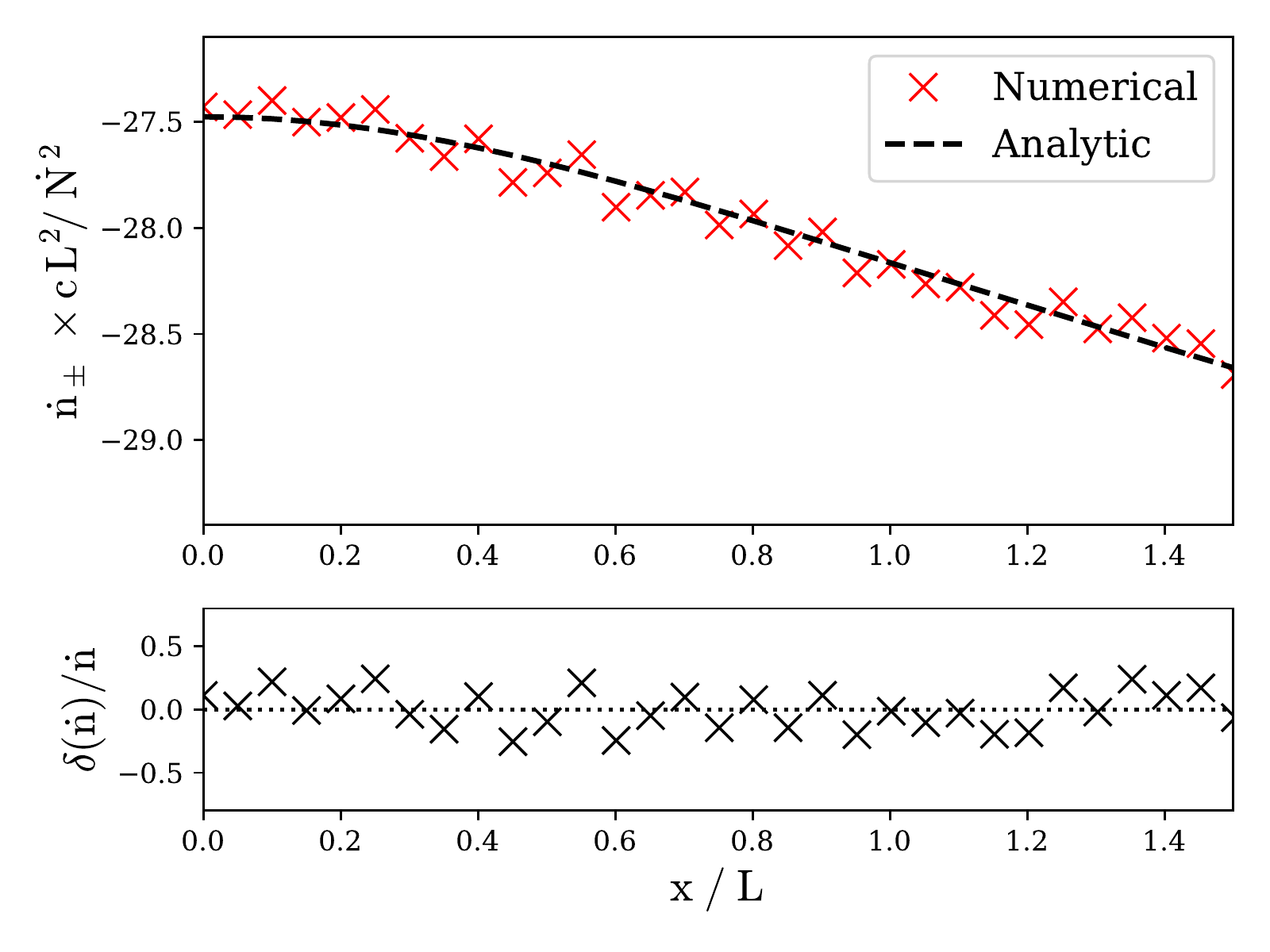}
  \caption{Monochromatic emitter test problem for $N_s \approx 10^6$. Upper panel: numerical (red hashes) and analytic (black line) pair production rate densities for two monochromatic, isotropic emitters with source separation $2\mathrm{L}$, evaluated as a function of radius in the plane normal to and  bisecting the line connecting the emitters. Lower panel: fractional difference between numerical and analytic values.}
  \label{fig:twopoint-monochromatic} 
\end{center} \end{figure}

Our procedure is as follows.  We generate a detailed sample of the radiation field using a Monte Carlo step that re-simulates the radiative transport and includes bremsstrahlung emission.\footnote{We use a ``fast light'' approximation that neglects the light-crossing time and allows us to avoid coupling snapshots of the plasma state taken at different coordinate times.} In this scheme, each radiation field sample $i$ is assigned a weight $w_i$ equal to the number of physical photons in the sample multiplied by a constant that is inversely proportional to photon sampling cadence.  Given a list of radiation field samples within a cell of coordinate volume $\Delta^3 x$, 
\begin{equation}
\label{eq:ndotnumerical}
\dot{n}_\pm \approx\dfrac{1}{2} \sum\limits_{i,j} \dfrac{w_i}{\sqrt{-g} \Delta^3 x} \dfrac{w_j}{\sqrt{-g} \Delta^3 x}  \dfrac{\epsilon^2}{k_i^0 k_j^0} \, \sigma_{\gamma\gamma} \, c ,
\end{equation}
where the Latin indices label radiation field samples in that cell.

Although Equation~(\ref{eq:ndotnumerical}) can be summed pairwise over all $n$ samples, it is more efficient to sample the sum over a subset of $m<n^2$ pairs $(i,j)$.  We set an upper limit on the number of pairs to consider and use reservoir sampling to obtain an unbiased subset from the full list.  In the limit that the $m$ is large, the error in Equation~(\ref{eq:ndotnumerical}) exhibits the usual $m^{-1/2}$ Monte Carlo scaling.

\subsection{Test Problems}

We now consider two tests to verify our method.  Both tests comprise two steady, isotropic, point-like photon sources in flat space separated by a distance $2L$. For each case, we measure the pair production rate along the perpendicular bisector of the line connecting the two sources (see Figure~\ref{fig:twopoint-geometry}).  

\begin{figure}[!ht] \begin{center}
  \includegraphics[width=\linewidth]{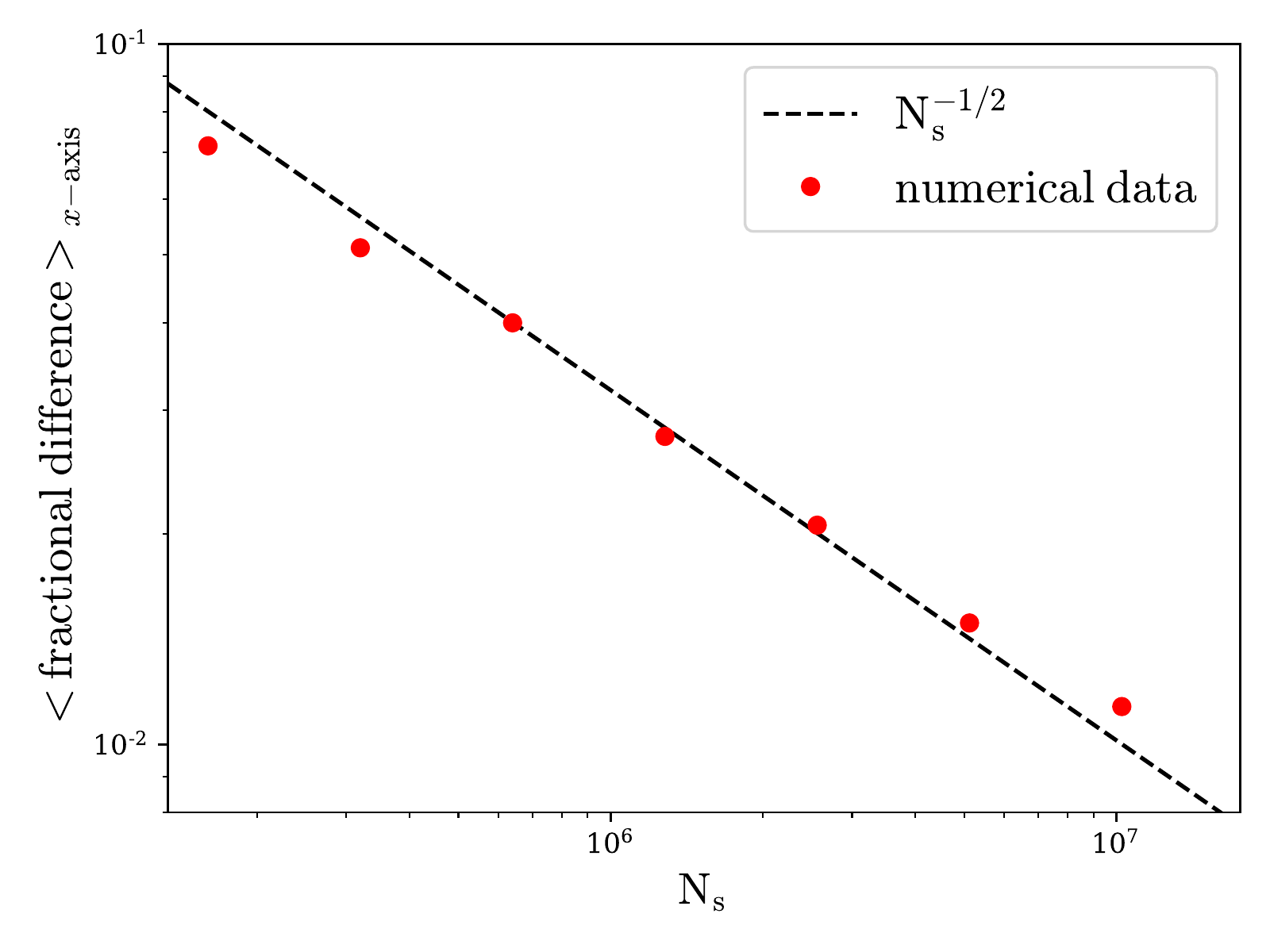}
  \caption{Code convergence. Averaged fractional difference between numerical and analytic pair production rate densities for the two-point, monochromatic, isotropic emitter problem as a function of number of field samples generated. The error scales $ \propto N_s^{1/2}$ as expected.}
  \label{fig:twopoint-monochromatic-convergence}
\end{center} \end{figure}

\subsubsection{Monochromatic Point Sources}

In the first test, each point source is monochromatic, and the pair production rate can be evaluated analytically (see \citealt{moscibrodzka11}; the test is provided here as a consistency check).  The center of mass energy $\epsilon$ is 
\begin{equation}
\epsilon^2 = - \dfrac{k_\mu k'^\mu}{2} = \dfrac{k^0 k'^0 \left(1 - \cos\theta \right) }{2},
\end{equation}
and 
\begin{equation}
\dfrac{\ud N_{\gamma}}{\ud^3 x} = \dfrac{\dot{N}_\gamma \sin^2\left(\theta/2\right)}{4 \pi L^2 c},
\end{equation}
where $\dot{N}_\gamma$ is the rate of isotropic photon production in the frame of the emitters. Then
\begin{equation}
\dot{n}_\pm(\theta) = \left( \dfrac{ \dot{N}_\gamma \sin^2\theta}{4 \pi L^2 c} \right)^2 \left( 1 - \cos \theta \right) \sigma_{\gamma\gamma}(\epsilon) \, c .
\end{equation}

Figure~\ref{fig:twopoint-monochromatic} compares $\dot{n}_\pm(x)$ computed analytic versus numerically in the upper panel and shows the fractional difference between the two evaluations in the lower panel.  Figure~\ref{fig:twopoint-monochromatic-convergence} shows the fractional difference between the domain-averaged numerical values and the analytic expression as a function of $N_s$, the number of samples of the radiation field. As expected, the error scales as $N_s^{-1/2}$.

\subsubsection{Power Law Spectrum Point Sources}
\label{section:powerlawspec}

In the second problem, we endow each point source with a power law spectrum with index $\alpha$ and cutoff frequencies $\nu_{\mathrm{min}} \ll \nu_e$ and $\nu_{\mathrm{max}} \gg \nu_e$, where $\nu_e \equiv m_e c^2/h$. In particular,
\begin{equation}
L_\nu = \left\{\begin{array}{ll} \quad \frac{L_0}{\nu_e} \, \left(\frac{\nu}{\nu_e}\right)^{\alpha} & \qquad \nu_{\mathrm{min}} < \nu < \nu_{\mathrm{max}} \\ \quad0 & \qquad \mathrm{otherwise} . \end{array}\right.
\end{equation}

\begin{figure}[!ht] \begin{center}
  \includegraphics[width=\linewidth]{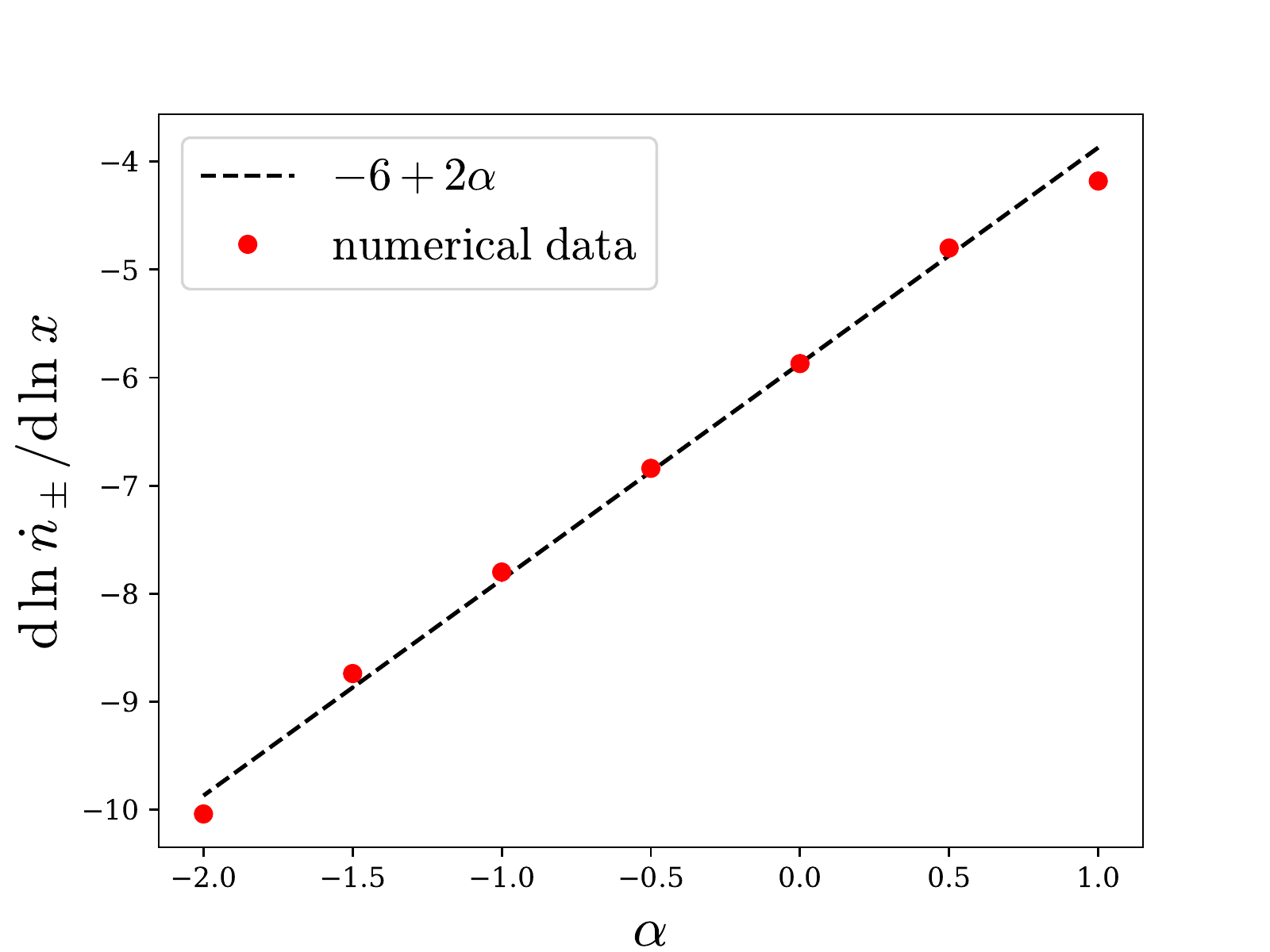}
  \caption{
  Radial dependence of pair production rate density versus source spectrum index. 
  Here, $\alpha$ is the index of the source radiation spectrum $L_\nu \sim \nu^\alpha$, and the slope $\ud \ln \dot{n}_\pm / \ud \ln x$ describes the asymptotic radial power law dependence of the pair production rate density versus distance $x$ from the source. The numerical results are plotted against and agree with the analytic estimate.
  }
  \label{fig:twopoint-spectrum-slopes}
\end{center} \end{figure}

The pair production cross section peaks for $\epsilon \sim 1$, and so the dominant contribution from photons with $\nu_{\mathrm{min}} \ll \nu_e$ will be through their interactions with high energy photons at frequencies $\nu = \nu_e / \nu_{\mathrm{min}}$, provided $\alpha$ is not too large (otherwise most pair production is by photons with $\nu \sim \nu_{\mathrm max}$).  In the astrophysical settings of interest to us, $\nu_{e}/\nu_{\mathrm{min}} \gg \nu_{\mathrm{max}}/\nu_e$, so we can neglect the dependence of $\dot{n}_\pm$ on $\nu_{\mathrm{min}}$.

Analytic evaluation of the pair production rate density for this test is difficult because the pair production cross section depends on energy in a non-trivial way, but the asymptotic scaling with $x$ is easy to compute.  At each $\theta = 2 \arctan(L/x) \approx 2 L/x$ for $x \gg L$, the dominant contribution to the rate integral is at $\epsilon \sim 1$ or $\nu \sim (\nu_e^2/\nu') (4 x^2/(L^2))$.  The product of the distribution functions thus scales as $x^{-4 + 2\alpha}$ because of the relationship between $\nu$ and $\nu'$ and $k^0 k'^0 \sim \nu \nu' \sim x^2$, implying\footnote{Rigorously: take the cross section to be a $\delta$ function in $\epsilon$ and integrate over $\nu$.} that 
\begin{align}
\label{eq:ndot-spectrum-scaling}
\dot{n}_\pm \sim x^{-6 + 2\alpha}.
\end{align}
The radial dependence of the pair production rate is therefore a non-trivial function of the source spectral index measured at pair-producing energies. Figure~\ref{fig:twopoint-spectrum-slopes} compares the numerically evaluated pair production rate density to the analytic estimate.

\subsection{Goldreich--Julian Charge Density}

There may be regions in a black hole magnetosphere where the charge density is insufficient to screen electric fields in the frame of the plasma and thus where the ideal MHD condition is violated.  In these regions, the unscreened electric field can accelerate electrons and positrons to sufficiently high energies that they produce photons above the pair production threshold.  Once this new generation of electron--positron pairs is produced, they themselves are accelerated in the unscreened electric field. This process can repeat over multiple generations, and ultimately, the \emph{pair cascade} will continue until enough charge has been produced to short out the potential (see, e.g., \citealt{sturrock1971,ruderman1975} and \citealt{beskin92} in the context of black holes).

The minimum charge density required to screen electric fields is known as the Goldreich--Julian charge density $n_{\mathrm{GJ}}$ \citep{goldreich69}.
In covariant language, a charge density $\rho_q$ is given by $\rho_q = -u^\mu j_\mu$, where $j^\mu$ is the four-current and $u^\mu$ is the four-velocity of the frame in which the charge density is measured; $\rho_q$ is thus a frame-dependent quantity. The four-current is \emph{always} given by Maxwell's equations $j^\mu = {F^{\mu\nu}}_{;\nu}$ (notice that the covariant derivative includes time derivatives).\footnote{When calculating $j^\mu$ from a GRMHD simulation, we first construct $F^{\mu\nu}$ from the four-velocity $u^\mu$, the magnetic induction four-vector $b^\mu$, and the ideal MHD condition $F^{\mu\nu}u_\nu = 0$ according to $F^{\mu\nu} = \epsilon^{\mu\nu\alpha\beta} u_\alpha b_\beta$, where $\epsilon^{\mu\nu\alpha\beta}$ is the Levi--Civita tensor. The four-current can then be computed from the inhomogeneous Maxwell equations, $j^\mu = {F^{\mu\nu}}_{;\nu}$. We use a finite-difference method across neighboring simulation locations and time slices to evaluate the derivative.} Goldreich and Julian's calculation is done in flat space, and the charge density is measured in the nonrotating frame. In our case, the choice of frame is less obvious. If the magnetosphere solution can be described by ideal MHD, then the uniquely sensible choice of frame is the fluid frame. In the Blandford--Znajek solution, however, there is no unique four-velocity associated with the force-free solution. Instead, we evaluate the charge density---which we will call the Goldreich--Julian density---in the normal observer frame:
\begin{equation}
\label{eq:ngj-defn}
n_{\mathrm{GJ}} e = - n^\mu j_\mu .
\end{equation}
In the normal observer frame $u_\mu = n_\mu \propto \left( 1, 0, 0, 0 \right)$ and using $j^\mu$ as the four-current of the \citet{blandford77}, hereafter BZ, split monopole solution, 
\begin{equation}
n_{\mathrm{GJ}} \approx \left( \dfrac{\bhspin B^r c^2}{ 4\pi G M e} \right) \dfrac{\left(1 + 2/x\right)^{1/2} \cos \theta}{x^3},
\end{equation}
where $x \equiv r/\mathcal{L}$, $\mathcal{L} \equiv GM/c^2$ and $B^r$ is now the radial component of the magnetic field at $x=1$ in spherical Kerr-Schild coordinates \citep[see][]{moscibrodzka11}.   

Assuming that the magnetic pressure $\sim$ the gas pressure and that both are of order $\rho c^2$, then for $B^r \approx 10^4 \sqrt{\dot{m}/m_8}$, where $m_8 \equiv M / \left( 10^8 M_\odot \right)$. The charge number density for $\bhspin = 0.94$ is 
\begin{equation} \label{eq:ngj-mckinney-fit}
n_{\mathrm{GJ}} \simeq 2.0 \times 10^{-1} \; \dot{m}^{1/2} \, m_8^{-3/2} \, \mathrm{cm}^{-3}.
\end{equation}

Notice that the charge density does not necessarily vanish in MHD, since ${\bf E} = 0$ in the plasma frame does not imply $\nabla \cdot {\bf E} \ne 0$.  Nevertheless, the MHD solution cannot be self-consistent where $n < n_{GJ}\, (= -u^\mu j_\mu /e)$, since Maxwell's equations cannot be satisfied.  Also note that if $n \gg n_{GJ}$, then it is not clear how to produce macroscopic regions with unscreened electric fields (gaps) in MHD unless there is an unresolved process that drives the number density towards zero.  
Finally, note that our ideal MHD simulations can never represent macroscopic regions where $\bE \cdot \bB \ne 0$, so they are incapable of recovering the dynamics of gaps and pair cascades.

\section{Spatial Distribution of Pair Production}
\label{sec:spatial}

We now provide a simple geometric treatment to motivate the function form of the drizzle pair production rate density due to each component of the background radiation spectrum. In the following section, we will use the numerical results to fit the model parameters. In general, the drizzle pair production rate density may be a function of time and space, and it can depend on model parameters like black hole spin.  Because of symmetries in the spacetime, however, we expect the \emph{mean} rate density to be independent of time and azimuth.

The density of pair-producing photons is
\begin{equation}
n_\gamma = n_{\gamma,\mathrm{synch}} + n_{\gamma,\mathrm{Compt}} + n_{\gamma,\mathrm{brems}},
\end{equation}
where the terms represent photons produced by direct synchrotron emission, Compton scattering, and direct bremsstrahlung emission respectively. In our models, scattered bremsstrahlung photons and direct synchrotron photons are both negligible near the pair-production threshold. 

Since $n_{\gamma,\mathrm{Compt}}$ and $n_{\gamma,\mathrm{brems}}$ have different spatial distributions, we neglect the $n_{\gamma,\mathrm{Compt}} n_{\gamma,\mathrm{brems}}$ cross-term, which is negligible compared to $ n_{\gamma,\mathrm{Compt}}^2$ and $n_{\gamma,\mathrm{brems}}^2$ over the bulk of the domain. We thus approximate the total drizzle pair production rate density as a sum of two independent terms due to self-interaction of Compton and bremsstrahlung photons respectively:
\begin{equation}
\dot{n}_\pm(r,\mu) \approx \dot{n}_{\pm,\mathrm{Compt}} + \dot{n}_{\pm,\mathrm{brems}}.
\end{equation}

\subsection{Compton Contribution}
\label{sec:sub-compton-contribution}

\setcounter{table}{0}
\begin{deluxetable*}{ cccccccccc }
\label{table:modelparameters}
\tablecaption{Time-Averaged RadGRMHD Model Parameters} 
\tablehead{ 
\colhead{model} & 
\colhead{$\bhspin$} &
\colhead{$m_8$} &
\colhead{$\dot{m}$} &
\colhead{$L_{\rm bol} / L_{\rm Edd}$} &
\colhead{$\epsilon_{\rm rad}$} &
\colhead{$L_{\pm} / \left( L_{\mathrm{BZ}} \Gamma_j\right)$} &
\colhead{$l_c$} &
\colhead{notes}
}
\startdata
A5 & $0.5$ & $33$ & $2.2 \times 10^{-5}$ & $4.7 \times 10^{-6}$ & $0.021$ & $1.0 \times 10^{-5}$ & $0.1$ & \multirow{4}{*}{M87-like} \\
A9 & $0.94$ & $33$ & $8.2 \times 10^{-6}$ & $1.5 \times 10^{-6}$ & $0.018$ & $3.9 \times 10^{-6}$ & $0.03$ &  \\
B5 & $0.5$ & $62$ & $9.2 \times 10^{-6}$ & $7.1 \times 10^{-7}$ & $7.7 \times 10^{-3}$ & $6.0 \times 10^{-7}$ & $0.02$ &   \\
B9 & $0.94$ & $62$ & $5.2 \times 10^{-6}$ & $5.6 \times 10^{-7}$ & $1.1 \times 10^{-3}$ & $6.7 \times 10^{-7}$ & $0.01$ &  \\ \hline
C & $0.5$ & $1$ & $1.1 \times 10^{-5}$ & $7.1 \times 10^{-7}$ & $6.5 \times 10^{-3}$ & $1.8 \times 10^{-7}$ & $0.02$ & \multirow{4}{*}{---} \\
D & $0.5$ & $1$ & $1.0 \times 10^{-6}$ & $1.3 \times 10^{-8}$ & $1.3 \times 10^{-3}$ & $4.6 \times 10^{-9}$ & $3 \times 10^{-4}$ & \\
E & $0.5$ & $1$ & $1.3 \times 10^{-7}$ & $2.9 \times 10^{-10}$ & $2.2 \times 10^{-4}$ & $2.7 \times 10^{-15}$ & $7 \times 10^{-6}$ & \\
F & $0.5$ & $1$ & $1.2 \times 10^{-8}$ & $4.2 \times 10^{-12}$ & $3.5 \times 10^{-5}$ &  $3.4 \times 10^{-18}$ & $1 \times 10^{-7}$ & \\
\enddata
\tablecomments{
From left to right: model name, dimensionless black hole spin parameter $\bhspin$, $m_8\equiv$ black hole mass in units of $10^8 M_\odot$, $\dot{m}\equiv$ black hole accretion rate in units of Eddington mass accretion rate $\dot{M}_{\mathrm{Edd}} = 2.22 \, m_8\  M_\odot \, \mathrm{yr}^{-1}$, time-averaged ratio of bolometric luminosity to Eddington luminosity, time-averaged radiative efficiency $\epsilon_{\rm rad} = L_{\rm bol} \, \dot{M}^{-1} c^{-2}$,
ratio of rest-mass pair luminosity to BZ luminosity, and compactness parameter (related to efficiency of pair production, see Section~\ref{sec:self-consistency} and Equation~(\ref{eq:compactness})). The values reported in this table include the bremsstrahlung contribution and thus differ from previous results.
}
\end{deluxetable*}

The pair production rate density is a strongly decreasing function of distance from the photon source.  Compton upscattered pair-producing photons come from regions of high electron temperature, and so the Compton contribution is likely to correlate strongly with regions of peak electron temperature, which exist both in the jet--disk boundary layer (see the \citealt{wong2020b} companion paper for a study of the jet--disk boundary layer) and close to the event horizon. Thus, the Compton contribution may be written as
\begin{align}
\label{eq:ndot-fit-compt}
& \dot{n}_{\pm,\mathrm{Compt}}(r,\mu) = \mathcal{A} \, \left(\dfrac{r}{\mathcal{L}}\right)^{-\alpha} \left(  e^{ - \mu^2/2\sigma^2 } + \right. \nonumber \\ 
& \qquad \left. \mathcal{B} \left( e^{-\left(\mu-\mu_f\right)^2/2\sigma_f^2} + e^{-\left(\mu+\mu_f\right)^2/2\sigma_f^2  } \right) \right)
\end{align}
where $\mathcal{A}$ is an overall normalization with dimensions of rate density, $\mathcal{B}$ describes the relative importance of the jet--disk boundary versus the midplane, $\mu \equiv \cos\theta$, $\sigma$ and $\sigma_f$ describe the scale heights of pair production in the disk and the boundary layer respectively, and $\mu_f$ is the location of the boundary layer. Following \moscref, we parameterize
\begin{equation}
\label{eq:funnel-wall-prescription}
{\mu_f}^2 = \dfrac{r+a}{r+b}.
\end{equation}
This model has $7$ parameters, $\mathcal{A}, \alpha, \mathcal{B}, \sigma, \sigma_f, a$, and $b$; however, we will find that the last four parameters can be fixed.

\subsection{Bremsstrahlung Contribution}

Bremsstrahlung photons near the pair-production threshold are emitted primarily in regions where the dimensionless electron temperature $\Theta_e \equiv k_B T_e/m_e c^2 \gtrsim \nicefrac{1}{2}$ (here $k_B$ is Boltzmann's constant). In our models, this region extends out to approximately $r = 10\ G M / c^2 \equiv r_{\mathrm{crit}}$ and corresponds physically to the domain in which viscous heating, electron cooling, and Coulomb cooling are in approximate balance. Because bremsstrahlung emissivity depends only on $n_i$, $n_e$, and $\Theta_e$, the geometry of the bremsstrahlung-driven region of pair production varies little from model to model.

Foresight from the numerical simulations and the presence of a radial cutoff at $r_{\mathrm{crit}}$ suggest that
\begin{align}
\label{eq:ndot-fit-brems}
&\dot{n}_{\pm,\mathrm{brems}}(r,\mu) = \nonumber \\ &\quad\mathcal{C} \, \dfrac{\zeta^2}{\zeta^2 + \mu^2} \left\{ \begin{array}{ll} \left(\dfrac{r}{\mathcal{L}}\right)^{-\kappa_1} & \quad  r < r_{\mathrm{crit}} \\ \left(\dfrac{r}{\mathcal{L}}\right)^{-\kappa_2} \; r_{\mathrm{crit}}^{\kappa_2-\kappa_1} & \quad r_{\mathrm{crit}} \le r, \end{array} \right.
\end{align}
where $\zeta$ parameterizes the dependence of the distribution on elevation $\mu$. This model has five parameters, $\mathcal{C}, \kappa_1, \kappa_2, r_{crit},$ and $\zeta$.  We find that fixing $\zeta = 0.4$ globally does not affect the quality of the fit.

The characteristics of the fluid and radiation field determine $\kappa_1$.  The bremsstrahlung spectrum just  above its peak follows a power law according to the behavior of $\Theta_e$ in the domain of emission, with $\alpha \approx -1$.  Using  Equation~(\ref{eq:ndot-spectrum-scaling}), we therefore expect $\kappa_2 = 6 + 2 \times 1 = 8$.

\section{Numerical Results}
\label{sec:results}

We now describe the results of our numerical pair drizzle simulations and provide fits for the model parameters described in \S~\ref{sec:spatial}.

\subsection{Simulation Parameters}

We consider the eight SANE radGRMHD models listed in Table~\ref{table:modelparameters}. The first four models have  M87-like parameters \citep{ryan18} with varying spin and mass. The second four models increase $\dot{m}$ at fixed mass and spin until radiative cooling becomes important \citep{ryan17a}.

Initial conditions for the fluid were produced by axi-symmetrizing three-dimensional nonradiative GRMHD models, which are less computationally expensive to evolve. 
In mapping from three to two dimensions, the no-monopoles constraint was enforced by computing $B^i$ from an axisymmetrized vector potential calculated from the original GRMHD simulations.

All fluid calculations were carried out in the modified Kerr-Schild (MKS) coordinates of \cite{mckinney04} with $h = 0.3$. The inner boundary was located within the event horizon, and the outer boundary was set at $r=200\, GM/c^2$.  The simulations were run at a resolution of $388$ radial zones by $256$ elevation zones.

\subsection{Simulation Outcomes}

\begin{figure}[!ht] \begin{center}
  \includegraphics[width=\linewidth]{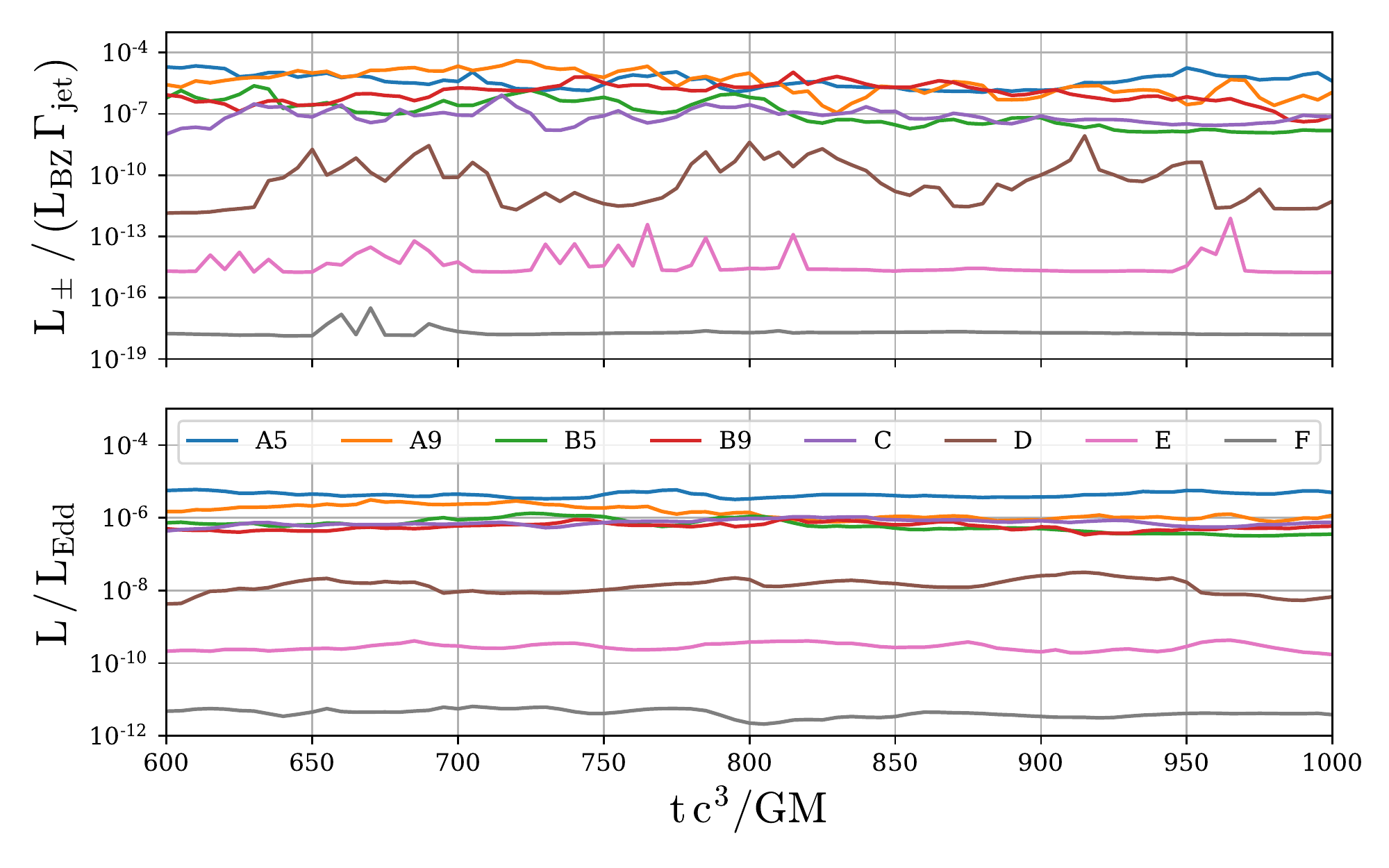}
  \caption{Time series of pair production rate and luminosity. Top: rest-mass pair drizzle luminosity divided by BZ jet power. Bottom: numerically calculated bolometric luminosity versus time. Over our range of models, time variability increase with $\dot{m}$ because the increasingly important Compton contribution scales more favorably than the bremsstrahlung one. }
  \label{fig:genvstime}
\end{center} \end{figure}

In our models the cumulative pair production rate fluctuates over four orders of magnitude, with both the domain-integrated and position-dependent pair production rates varying on timescales as short as the fluid dump cadence $5\,M$.  Following the discussion in  Section~\ref{sec:sub-compton-contribution}, we find that the time- and azimuth- averaged $\dot{n}_{\pm,\mathrm{Compt}}$ peaks in the midplane and in {\textit{hotspot}} regions (characterized by high $\Theta_e$) that lie within the jet--disk boundary layer. The structure and locations of the hotspot regions are highly variable. 

The pair drizzle luminosity is
\begin{equation}
\label{eq:drizzleluminosity}
L_{\pm} \equiv 2 m_e c^2 \, \dot{N}_\pm \, \Gamma_{\mathrm{jet}},
\end{equation}
where $\Gamma_{\mathrm{jet}}$ is the bulk Lorentz factor at large $r$ and the domain-integrated pair production rate is $\dot{N}_\pm \equiv \int \sqrt{-g}\, \ud^3 x\, \dot{n}_\pm$. Figure~\ref{fig:genvstime} shows the time variability in both the pair drizzle luminosity and the background bolometric luminosity. Although both quantities exhibit variations, fluctuations in the former occur on shorter timescales and with greater amplitude. These variations are primarily caused by transient hotspot regions associated with plasmoids that form within the jet--disk boundary layer and travel across the domain. For models in which bremsstrahlung is the primary source of photons near the pair production threshold, variability is decreased. This is particularly evident in model F. For models in which Compton upscattering is the primary source of photons near the pair production threshold, time variability decreases as $\dot{N}_\pm$ increases. This is unsurprising, since increasing the pair production rate requires a larger fraction of the domain to be in a steadily pair-producing regime.

\begin{figure}[!ht] \begin{center}
  \includegraphics[width=\linewidth]{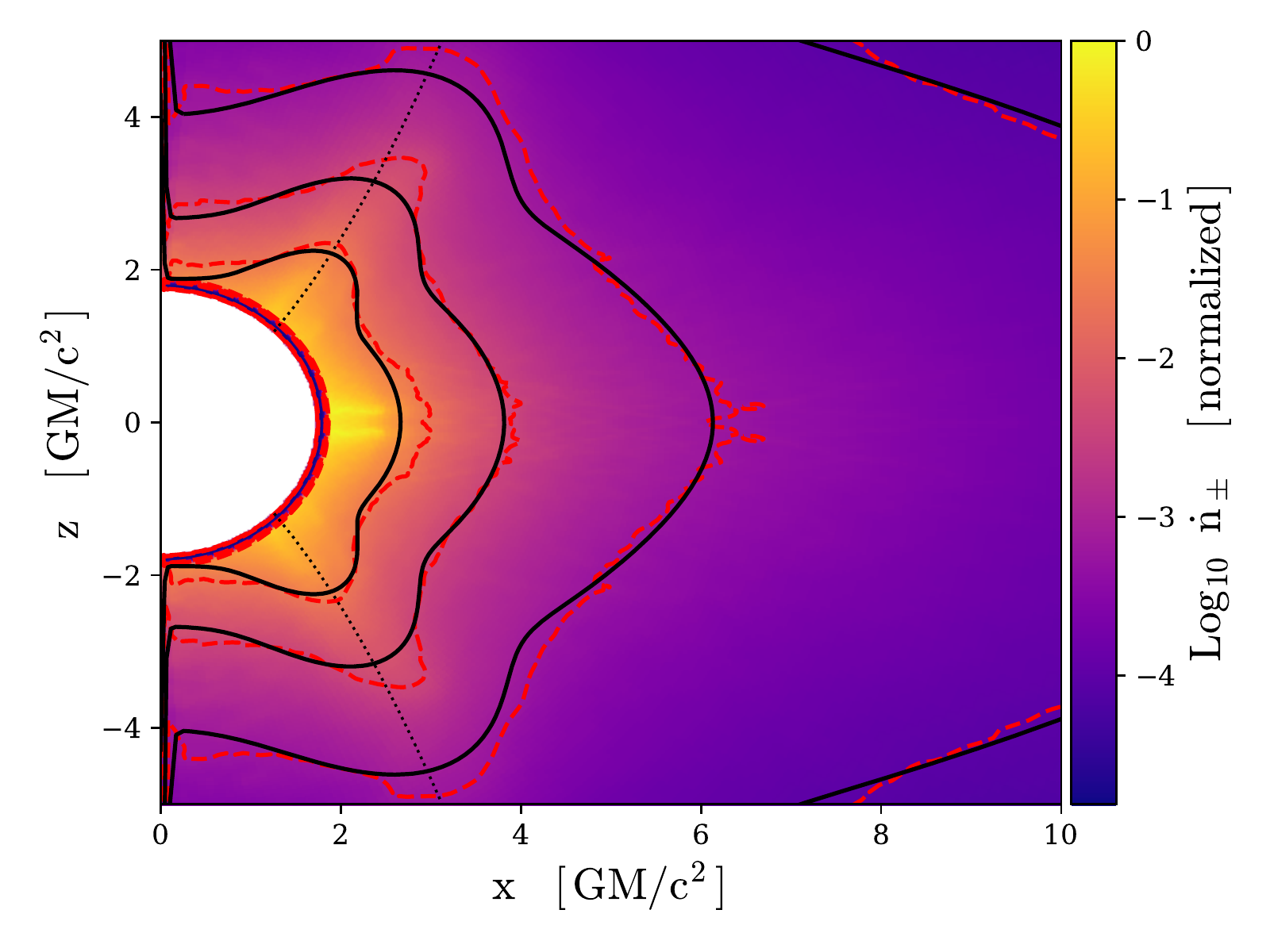}
  \caption{Pair production rate density (Model C). Numerically evaluated, time-averaged pair production rate density $\dot{n}_\pm$ as a function of position over domain for model C after vertical symmetrization over the disk midplane. Horizontal axis shows radial coordinate and vertical axis shows height above midplane. Solid colors correspond to $\log_{10}(\dot{n}_\pm)$. Dashed red lines track contours in numerical value and solid black lines represent contours of model with fit parameters.}
  \label{fig:contour-fit}
\end{center} \end{figure}

In fitting the time- and azimuth- averaged pair production rate, we must fit the location of the boundary layer.  Various techniques for defining and tracking the extent of the jet have been explored in the literature (e.g., \citealt{narayan12, yuan15, moscibrodzka16}).  We find that fitting the location of the jet--disk boundary for each model is not justified by the improvement in fit to the pair production rate, and we simply fix $a = \nicefrac{1}{2} \,\mathcal{L}$ and $b= 3\, \mathcal{L}$ in Equation~(\ref{eq:funnel-wall-prescription}).  Similarly we use $\sigma = 0.5$ and $\sigma_f = 0.1$ in Equation~(\ref{eq:ndot-fit-compt}) for all models.  This finding is unsurprising because the scale height of the disk and the width of the jet are independent of $\dot{m}$ and $m_8$ in our model set.  

The parameters are likely to depend on magnetization, but we cannot evaluate this dependence because we consider only SANE models. The parameters may also depend on numerical resolution and the dimensionality of the model, which we also cannot evaluate with the existing model set.  It is possible, for example, that the peak in the jet--disk boundary layer could increase as resolution increases and dissipation is concentrated in a narrower region within the boundary.

Figure \ref{fig:contour-fit} shows the time-and-azimuth averaged pair production rate for model B with fitting function contours overplotted. The fit is more accurate for inner regions of the disk where the $\dot{n}_\pm$ is large, but the fit works well even at larger radius.

In the low-$\dot{m}$ regime, we find 
\begin{subequations}
\label{eq:parameter-values-base}
\begin{align}
\label{eq:parameter-values}
\mathcal{A}(m_8, \dot{m}) &\approx 5.7 \times 10^{30} \; \dot{m}^{5.8} \; m_8^{-1.4} \\
\mathcal{B}(m_8, \dot{m}) &\approx 5.4 \times 10^{-4} \; \dot{m}^{-4/5} \\
\mathcal{C}(m_8, \dot{m}) &\approx 1.9 \times 10^{15} \; \dot{m}^4 \; m_8^{-2} \\
\alpha &\approx 4.9 \; \dot{m}^{-0.04} \\
\kappa_1 &\approx 2 \\
\kappa_2 &\approx 8 .
\end{align}
\end{subequations}
Again in the low-$\dot{m}$ regime, the total pair creation rate, integrated over the entire simulation domain, is well fit by 
\begin{equation}
\label{eq:Ndot-fit}
\dot{N}_\pm(m_8,\dot{m}) = 3.4 \times 10^{64} \; \dot{m}^{5} \; m_8^{1.5}.
\end{equation}

Pairs are born with a broad spectrum of energies. The Lorentz factor of each lepton $\gamma_{\mathrm{FF}}$ as measured in the plasma fluid frame $u_\mu$ can be computed from the $p^\mu$ of the interacting photons since momentum is conserved. In the jet, the average pair is created with $\gamma_{\mathrm{FF}} \approx 10$. This result is consistent with \moscref.

\section{Discussion}
\label{sec:discussion}

We have modeled drizzle pair production in simulations of SANE (low magnetic flux) black hole accretion flows in the mildly radiatively efficient regime and for select models corresponding to M87.  The accretion simulations we consider model electron thermodynamics and radiative processes. Our models differ from \moscref in several respects.  First, electron heating is treated using the \cite{howes10} model for dissipation at the bottom of a turbulent cascade.  This model partitions dissipation approximately equally between electrons and ions when $B^2/(8\pi) \gtrsim P_{\mathrm{gas}}$ and preferentially heats the ions otherwise.   Second, Coulomb coupling between ions and electrons is included.  This transfers energy from ions to cooler electrons and is a significant source of electron heating near the midplane at small radius \citep{ryan17a}.   Third, we self-consistently treat the transfer of momentum and energy between the plasma and the radiation field using the {\tt ebhlight} code \citep{ebhlight}.  Finally, we consider bremsstrahlung emission when estimating the pair production rate (but not in the radGRMHD simulation, where it is energetically sub-dominant).  Since bremsstrahlung produces a large population of photons with $h\nu \sim k T \sim 10 m_e c^2$, it can be important for pair production.   

Our results largely agree with the analysis presented in \moscref.  Still, there are interesting new questions we can answer.  First: motivated by a new understanding of M87 based on the EHT 2017 results, is M87 likely to have a charge-starved magnetosphere?  Second, are there differences in the geometry of pair production between state-of-the-art models and \moscref's more simplified treatment of electron thermodynamics?

\subsection{Drizzle versus Gaps}

Pair drizzle can prevent the black hole magnetosphere from becoming charge-starved and thereby forestall the opening of gaps and the generation of pair cascades. To see this, we compare the total number of available charges from both pairs and plasma $n_{\mathrm{avail}} \equiv n_\pm + n_{\mathrm{pl}}$ to the Goldreich--Julian charge density drawn from Equation (\ref{eq:ngj-mckinney-fit}), which uses the normal observer frame in its calculation.  Using the characteristic time $\mathcal{T} \equiv \mathcal{L}/c$, we set $n_\pm \approx \dot{n}_\pm \mathcal{T}$.  At $r = 2 G M/c^2$ and along the pole at $\theta = 0$ (where $n_{\mathrm{avail}}$ is small since $n_{\mathrm{pl}}$ is negligible), the ratio is
\begin{align}
\label{eq:ngj-ratio}
\dfrac{n_{\mathrm{avail}}}{n_{\mathrm{GJ}}} &\; \approx 4.4 \times 10^{17} \; \dot{m}^{7/2} \ m_8^{3/2} \, \times \nonumber \\
& \; \left( 1 + 1.1 \times 10^{12} \; \dot{m}^{6/5} + 9.9 \times 10^{14} \; \dot{m}^2 \right).
\end{align}
The terms in parentheses correspond to pair production by bremsstrahlung, Comptonized photons from the jet--disk boundary, and Comptonized photons from the midplane.  Evidently for $10^{-10} < \dot{m} < 10^{-4}$ the Comptonized boundary photons dominate; for $\dot{m} \gtrsim 10^{-7}$ the midplane photons dominate bremsstrahlung; and the midplane become increasingly important as $\dot{m}$ increases.  The midplane may be more important than the boundary as $\dot{m} \gtrsim 10^{-3.7}$, but that extrapolates beyond the range of validity of our models.

\begin{figure}[!ht] \begin{center}
  \includegraphics[width=0.9\linewidth]{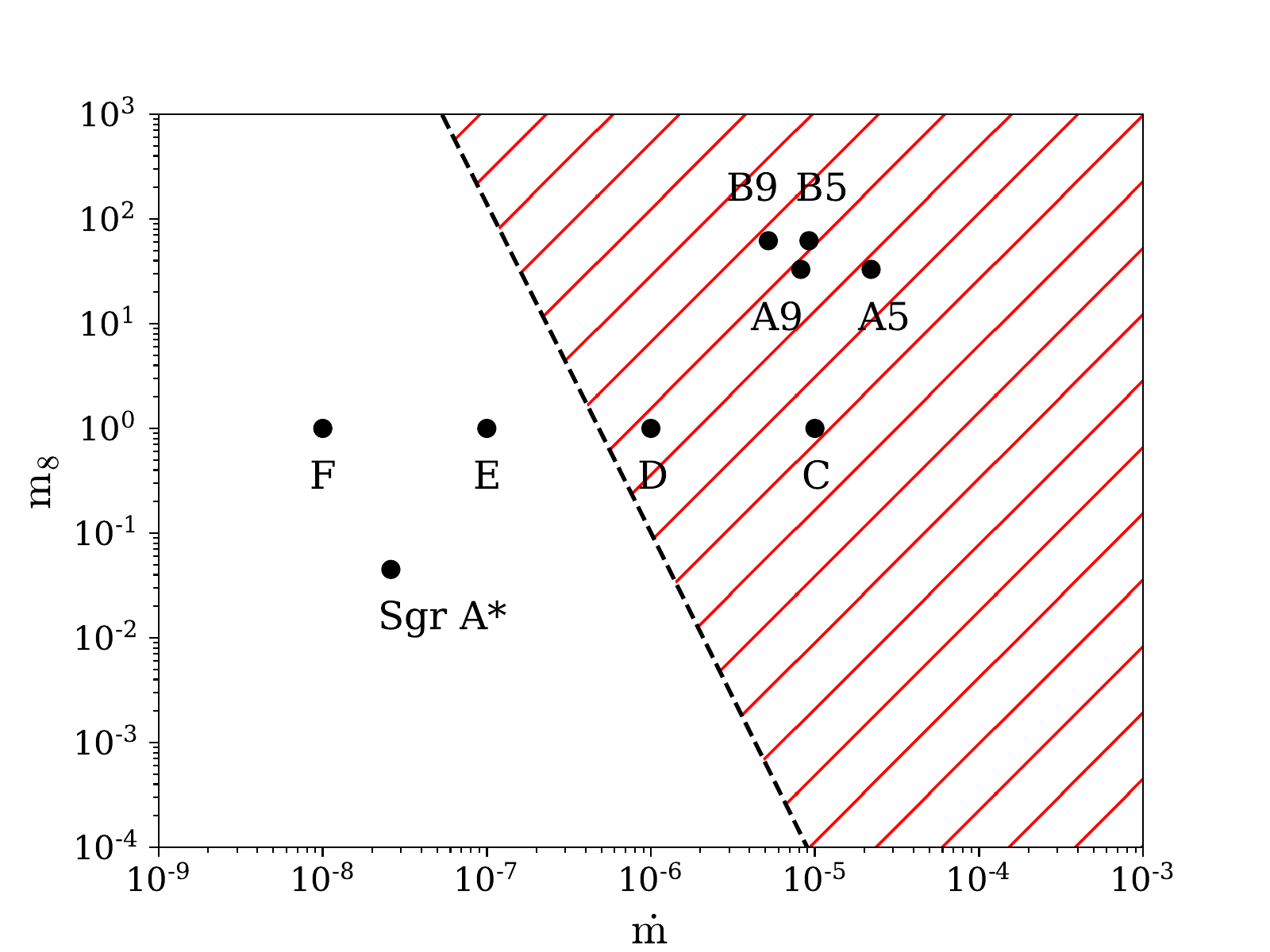}
  \caption{
  SANE $\bhspin = 0.5$ and $\bhspin = 0.94$ models in the $\dot{m}, m_8$ plane.  The red hash marks show regions where the ratio of Goldreich--Julian density to radiation-GRMHD number density is below unity.  In the unhatched region the MHD approximation is not self-consistent.   
  }
  \label{fig:ngj-ratio}
\end{center} \end{figure}

\begin{figure}[!ht] 
\begin{center}
  \includegraphics[width=\linewidth]{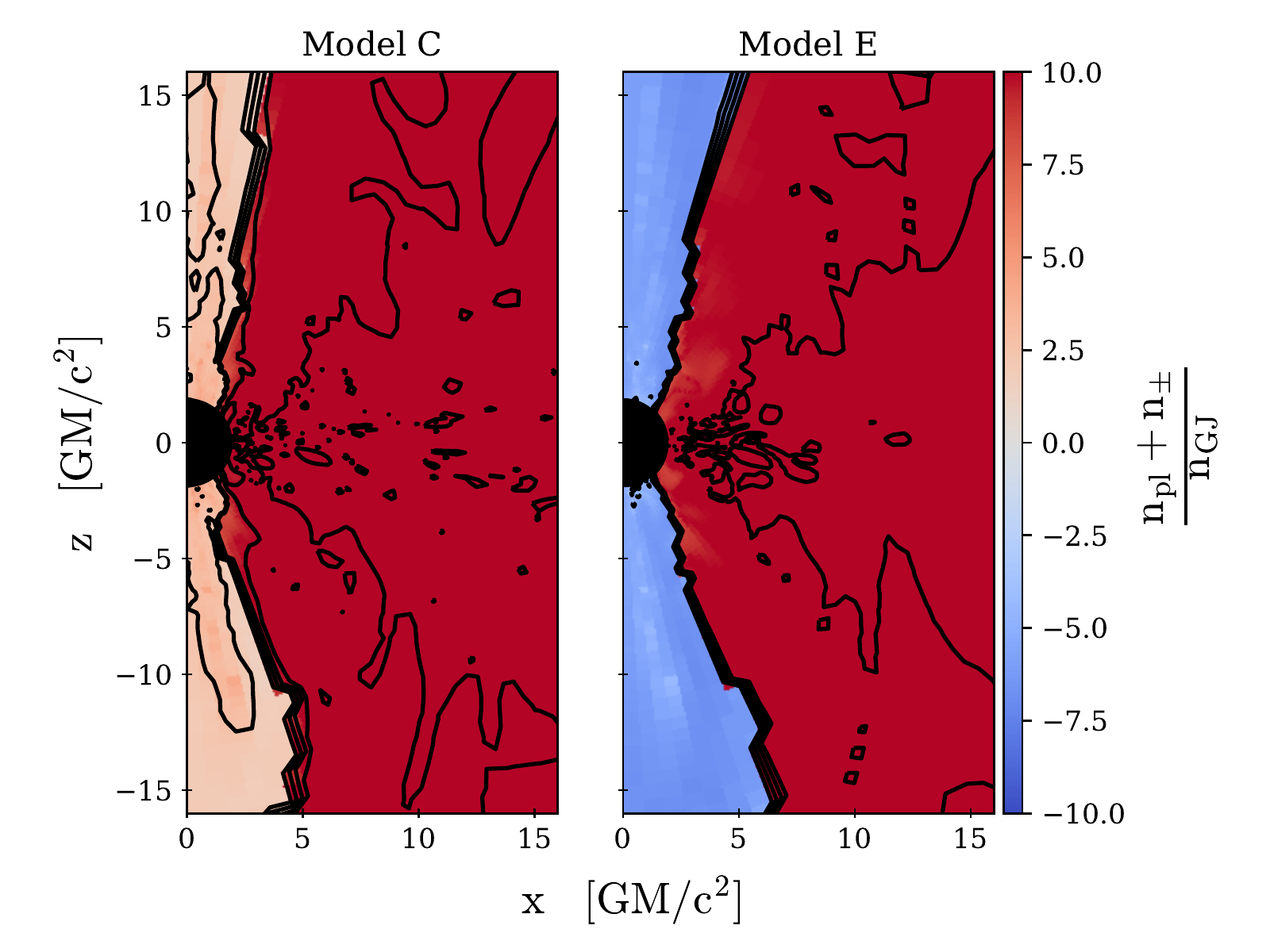}
  \caption{
  Ratio of available charge to Goldreich--Julian density (Equation~(\ref{eq:ngj-ratio})) for Models C and E ($\bhspin = 0.5, m_8 = 1$ with $\dot{m} = 1.1 \times 10^{-5}$ and $1.3 \times 10^{-7}$ respectively). Black contours are evenly spaced in the log of the ratio. The black circle is the event horizon.  Evidently the ratio is well above unity in the disk in both models, while the ratio in model E in the jet is far below unity and the MHD approximation is not self-consistent.  Although pair cascades are not included in our model, they would appear difficult to initiate anywhere in model C and likely in the jet region of model E.   
  }
  \label{fig:ngjdomain-ratio}
\end{center} 
\end{figure}

In regions where $n_{\mathrm{avail}} / n_{\mathrm{GJ}} < 1$, the MHD approximation is not self-consistent.  It seems likely that the outcome is a pair cascade (although this is not computable in our model) that increases $n_{\rm avail}$ by drawing on the free energy of the electromagnetic field and the radiation field until $n_{\mathrm{avail}} / n_{\mathrm{GJ}} \sim 1$. Figure~\ref{fig:ngj-ratio} shows where in parameter space, according to Equation~(\ref{eq:ngj-ratio}), the MHD approximation is not self-consistent.  Figure~\ref{fig:ngjdomain-ratio} maps $n_{\mathrm{avail}} / n_{\mathrm{GJ}}$ in the poloidal plane for models C and E.

\subsection{Drizzle Pair Production Power}

As a black hole spins, it drags spacetime and the magnetic field lines near the horizon with it. These field lines produce an outward Poynting energy flux as they wind around the pole, via the Blandford--Znajek (BZ) mechanism \citep{blandford77}. The BZ mechanism is a favored explanation for the source of black hole jet power. The BZ luminosity is given by
\begin{equation}
L_{\mathrm{BZ}} \equiv \int\limits_{\mu^2 > \mu_f^2} T^r_{\substack{\,t\\\!\mathrm{EM}}}  \sqrt{-g} \ \ud \theta \, \ud \phi,
\end{equation}
where $T^r_{\substack{\,t\\\!\mathrm{EM}}} = b^2 u^r u_t - b^r b_t$ is the radial energy flux for the electromagnetic component of the stress-energy tensor.  We find that the numerically computed values of $L_{\mathrm{BZ}}$ for our simulations match the fit given by Equation (36) of \moscref,
\begin{equation}
\label{eq:lbz-model}
L_{\mathrm{BZ}} \approx 8 \times 10^{45} \left( 1 - \sqrt{1 - \bhspin^2}\right)^2 \dot{m}\, m_8 \;\; \mathrm{erg}\ \mathrm{s}^{-1},
\end{equation}
when $\bhspin = 0.5$.\footnote{Perturbative calculations of the BZ luminosity (to higher orders in the hole frequency) have been computed and compared to numerical simulation by, e.g., \citet{tanabe2008bzhigherorder} and \citet{tchekhovskoy2010bhspinagn}. \citet{tchekhovskoy11} provided a generalized formula similar to the one given by \moscref that also accounts for different magnetic fluxes near the horizon and thus treats both SANE \emph{and} MAD accretion states.}

We now ask what fraction of the jet power can be accounted for by drizzle pairs. Using Equations~(\ref{eq:drizzleluminosity}) and~(\ref{eq:Ndot-fit}),
\begin{equation} \label{eq:power-scaling}
\dfrac{L_{\pm}}{L_{\mathrm{BZ}} \Gamma_\mathrm{jet}} \approx 3.8 \times 10^{14} \; \dot{m}^{4} \, m_8^{1/2}.
\end{equation}
Scaling Equation~(\ref{eq:lbz-model}) to M87 using EHT results \citep{PaperV} and assuming that $\bhspin \simeq 0.5$, $\dot{m} \simeq 10^{-5}$, and $m = 6.5 \times 10^9$, then $L_{\pm} \simeq 3 \times 10^{36} \; \Gamma_\mathrm{jet}$.  In order for the drizzle pair luminosity to be comparable to M87's X-ray luminosity $\approx 10^{42} \;\mathrm{erg}\,\mathrm{sec}^{-1}$, the typical pair would have to be born with an exceedingly high Lorentz factor $\Gamma_\mathrm{jet} > 10^6$.  Thus, although drizzle-produced pairs may become important at $\dot{m} \sim 10^{-4}$ Eddington (suggested by Equation~(\ref{eq:power-scaling}), but outside our model space), they account for a small fraction of total BZ power.

\subsection{Variability and the Radiation Model}

Our fits for $\dot{n}_\pm$ and $\dot{N}_\pm$ represent the time- and azimuth- averaged behavior of the background $\gamma\gamma$ pair production process. In contrast, the instantaneous $\dot{n}_\pm$ does not peak along the entire boundary layer at once, but rather inside isolated island-like structures or plasmoids. The plasmoids are elongated in the radial direction, extend several M in width, and tend to travel along the boundary and evolve on timescales comparable with the dynamical time (see \citealt{Nathanail2020} and \citealt{Ripperda2020} for a discussion of plasmoids in nonradiative models). The plasmoid evolution depends strongly on the model parameters and the electron thermodynamics. 

The highly variable plasmoid emission is dominated by Comptonized synchrotron photons rather than bremsstrahlung, which is generated mainly in the midplane at large radius and is relatively steady. Thus, in general, the high variability we observe in the domain-integrated pair production rate $\dot{N}_\pm$ is due to the rapid evolution of the plasmoids. 

The scaling relations provided above have a limited range of validity.  For $\dot{m} \gtrsim 10^{-5}$, radiative cooling is strong enough to qualitatively change the electron temperature distribution and thus the distribution of pair-producing photons.  We are currently unable to explore this behavior because of the increasing computational intractability of running Monte Carlo radGRMHD simulations as optical depths to photon scattering increase and cooling times decrease relative to the light crossing time of the domain.

\subsection{Limitations and Self-Consistency}
\label{sec:self-consistency}

Because our model only considers pair production in a post-processing step, it cannot account for any back-reaction of drizzle pairs on the radiation field or underlying fluid dynamics.  To check the self-consistency of this approximation, we can estimate the compactness parameter~\citep[e.g.,][]{salvati83} 
\begin{equation}
l_c \equiv \dfrac{L_\gamma}{\mathcal{L}} \dfrac{\sigma_T}{m_e c^3},
\label{eq:compactness}
\end{equation}
which is proportional to the optical depth to pair production. When $l_c \gg 1$, pair production cannot be treated as a perturbative process.  In our models $l_c$ ranges from $10^{-7}$ to $10^{-1}$ (see Table~\ref{table:modelparameters}), so our treatment is self-consistent.

We assume that the electron distribution is purely thermal; however, since the plasma is everywhere collisionless in all our models (the Coulomb scattering mean free path is large compared to $G M/c^2$), the plasma need not fully relax to a thermal distribution.  Moreover there is evidence for nonthermal electrons in both observations of low $\dot{m}$ accreting black holes (e.g., NIR emission in the case of Sgr A*) and in simulations of collisionless, turbulent plasmas \citep[e.g.,][]{kunz16}.  The presence of non-thermal electrons in a high-energy tail  can result not only in higher-energy synchrotron photons, but also in an increase in Compton scattering events that increase photon energies to above the pair-producing threshold.

Our models used the \citet{howes10} prescription for heating due to dissipation, in which the electron heating is driven by a Landau-damped turbulent cascade process.  Other prescriptions \citep[e.g.,][]{rowan17, werner2018, kawazura19} would naturally produce a different electron temperature distribution. Because the pair luminosity depends strongly on electron temperature, modifications to the electron thermodynamics could significantly alter our results in ways that are difficult to assess without re-running the radGRMHD models.

We computed pair production rates in post-processing using the fast-light approximation, in which it is assumed that the fluid does not change appreciably over the time it takes for light to travel across the simulation domain. It is possible but computationally expensive to dispense with this approximation (\emph{slow-light}).  Performing a full slow light calculation would undoubtedly alter the pair production rate density on small length- and time-scales, but notice that the total radiative energy budget is conserved in both fast- and slow- light treatments, and so unless the fast-light approximation dramatically changes $\langle n_\gamma^2 \rangle/\langle n_\gamma \rangle^2$, the time-averaged pair production rate should not change significantly.

Finally, our models were limited to moderate resolution and two dimensions because of the computational expense of running full radGRMHD simulations. Increasing resolution and especially performing simulations in three dimensions could change the profile of the jet--disk boundary layer and alter the dynamics of the plasmoid hotspots that develop within it. Since drizzle pairs production peaks near the hotspots and is strongly dependent on the plasma temperature, the structure of $\dot{n}_\pm$ may change significantly with increased resolution or in the case of fully three-dimensional simulations.

\section{Summary}
\label{sec:summary}

We have modeled pair production due to the collision of photons in the background radiation field (here referred to as drizzle pair production) for sub-Eddington black hole accretion systems in the SANE state. Our plasma model is based on radiation-GRMHD simulations using the {\tt{}ebhlight} code \citep{ryan17a,ryan18}, which evolves the plasma and the full energy-dependent photon distribution.  The radGRMHD evolution includes synchrotron emission, absorption, and Compton scattering.  It also separately evolves ion and electron internal energies and explicitly accounts for dissipation using the electron heating prescription of \citet{howes10}.  We post-processed the fluid data using Monte Carlo radiation transport to track pair production due to photon-photon collisions. In the post-processing we included bremsstrahlung emission, which is a potentially important source of photons near the pair production threshold. 

Our approach closely follows \moscref and extends it in several ways.  We use energetically self-consistent radiation-GRMHD models rather than nonradiative GRMHD models.  We incorporate a dissipation model rather than fixing the ion-to-electron temperature ratio.  We study multiple black hole spins ($\bhspin = 0.5$ and $0.94$).   Finally, we include bremsstrahlung emission, which is non-negligible at frequencies near the pair-production threshold.   

\vspace{0.5em}

Our key findings are:

\vspace{0.5em}

1. The importance of cooling increases as accretion rate increases. This leads to a shallower dependence of the source-integrated pair production rate on $\dot{m}$ than in \moscref. \vspace{0.2em}

2. The spatial distribution of pair production peaks within the jet--disk boundary, in contrast to \moscref. This is because electron temperature peaks in the boundary layer, and drizzle pair production closely follows the electron temperature profile.  These results are summarized by Equations~(\ref{eq:ndot-fit-compt}),~(\ref{eq:ndot-fit-brems}), and~(\ref{eq:Ndot-fit}), with the parameter values reported in Equation~(\ref{eq:parameter-values-base}). \vspace{0.2em}

3. The pair production rate density can be divided into spatially distinct bremsstrahlung and Comptonized synchrotron components.  The bremsstrahlung component is comparatively steady and lies in the midplane, outside the midplane Comptonized component.  The bremsstrahlung component is weaker than the Comptonized component for all models considered here. \vspace{0.2em}

4. The Comptonized component from the midplane becomes comparatively larger as $\dot{m}$ increases, but it is dominated by the boundary layer Comptonized component for all models considered in this paper. \vspace{0.2em}

5. The drizzle pair production rate is time variable, with the difference between subsequent samples occasionally approaching four orders of magnitude. These variations are dominated by fluctuations in the synchrotron and Compton components of the background radiation field within the jet--disk boundary. \vspace{0.2em}

6. We confirm the finding of \moscref that the drizzle process (in M87-like SANE models) produces a background pair density that is far above the Goldreich--Julian density.  This suggests that it will be difficult to open gaps absent some dynamical process that is not incorporated in our models. \vspace{0.2em}
    
7. We confirm the finding of \moscref that drizzle pair production in Sgr A*-like SANE models is too feeble to keep the pair density above the Goldreich--Julian density.  In GRMHD models the expected Blandford--Znajek power is $\sim 2.8 \,\bhspin^2 (\phi/15)^2 \dot{M} c^2 = 1.6 \times 10^{38}\, \bhspin^2 (\phi/15)^2 (\dot{M}/(10^{-9} \msun \mathrm{yr}^{-1})\, \mathrm{erg}\, \mathrm{sec}^{-1}$.  The difficulty in firmly identifying a jet with comparable power suggests it is not present.  \cite{parfrey19} PIC-based magnetosphere model suggests that this is {\em not} due to a fundamental change in the BZ power for charge-starved magnetospheres.

\vspace{0.8em}

In future work we plan to explore drizzle pair production in MAD models where the increased electron temperatures and magnetic field strengths provide a more favorable environment for high-energy photons production.  We also plan to extend the calculations to three dimensions, which will provide the opportunity to study transient behavior associated with the characteristic nonaxisymmetry of MAD models.

\acknowledgements

The authors thank Ricardo Yarza, Sasha Philippov, 
Yajie Yuan, and Monika Mo\'scibrodzka for productive discussions. The authors also thank the anonymous referee for comments that improved the presentation and clarity of the text.
This work was supported by the National Science Foundation under grants AST 17-16327 and PIRE 1743747, by a Donald C.~and F.~Shirley Jones Fellowship to G.N.W., and by a Richard and Margaret Romano Professorial scholarship to C.F.G.

\bibliographystyle{aasjournal}
\bibliography{biblio}

\end{document}